\documentclass[twocolumn,aps,prl,floatfix,superscriptaddress]{revtex4-1}
\usepackage{amsmath,amssymb,amsfonts}
\usepackage{pifont}
\usepackage{graphicx,float}
\newcommand{\cmark}{\ding{51}}%
\newcommand{\xmark}{\ding{55}}%

\newcommand{\abs}[1]{\left| #1 \right|}
\newcommand{\bra}[1]{\left\langle #1 \right|}
\newcommand{\ket}[1]{\left| #1 \right\rangle}

\renewcommand{\epsilon}{\varepsilon}

\def\VR{\kern-\arraycolsep\strut\vrule &\kern-\arraycolsep}
\def\vr{\kern-\arraycolsep & \kern-\arraycolsep}
\usepackage{sistyle}
\usepackage{soul}
\sethlcolor{lightblue}

\begin{document}
\title{Entanglement: \emph{Quantum} or \emph{Classical}?}
\author{Dilip Paneru}
\affiliation{Department of Physics, University of Ottawa, 25 Templeton Street, Ottawa, Ontario, K1N 6N5 Canada}
\author{Eliahu Cohen}
\affiliation{Department of Physics, University of Ottawa, 25 Templeton Street, Ottawa, Ontario, K1N 6N5 Canada}
\affiliation{Faculty of Engineering and the Institute of Nanotechnology and Advanced
Materials, Bar Ilan University, Ramat Gan 5290002, Israel}
\author{Robert Fickler}
\affiliation{Department of Physics, University of Ottawa, 25 Templeton Street, Ottawa, Ontario, K1N 6N5 Canada}
\affiliation{Photonics Laboratory, Physics Unit, Tampere University, Tampere, FI-33720, Finland}
\author{Robert W. Boyd}
\affiliation{Department of Physics, University of Ottawa, 25 Templeton Street, Ottawa, Ontario, K1N 6N5 Canada}
\author{Ebrahim Karimi}
\email{ekarimi@uottawa.ca}
\affiliation{Department of Physics, University of Ottawa, 25 Templeton Street, Ottawa, Ontario, K1N 6N5 Canada}
\begin{abstract}
From its seemingly non-intuitive and puzzling nature, most evident in numerous EPR-like gedankenexperiments to its almost ubiquitous presence in quantum technologies, entanglement is at the heart of modern quantum physics. First introduced by Erwin Schr\"odinger nearly a century ago, entanglement has remained one of the most fascinating ideas that came out of quantum
mechanics. Here, we attempt to explain what makes entanglement fundamentally different from any classical phenomenon. To this end, we start with a historical overview of entanglement and discuss several hidden variables models that were conceived to provide a classical explanation and demystify quantum entanglement. We discuss some inequalities and bounds that are violated by quantum states thereby falsifying the existence of some of the classical hidden variables theories. We also discuss some exciting manifestations of entanglement, such as N00N states and the non-separable single particle states. We conclude by discussing some contemporary results regarding quantum correlations and present a future outlook for the research of quantum entanglement.
\end{abstract}
\maketitle

\section{Introduction}
Until the beginning of the twentieth century, classical mechanics, e.g. Newtonian or Lagrangian, together with Maxwell's electrodynamics were successful in describing and prognosticating nearly all physical phenomena. Eventually, classical physics failed to describe several effects such as black-body radiation, Compton scattering, and the photoelectric effect \cite{photoelektric}. 
Starting with the introduction of the apparently smallest energy quantum by Max Planck \cite{Planck}, the formalism of quantum mechanics was developed in the 1920s to describe the atomic and subatomic world. Since its inception, quantum theory has found numerous theoretical and practical applications in physics, and has even branched out to areas such as biology~\cite{bio1,bio2,bio3}, chemistry~\cite{chem1,chem2,chem3}, and computer science~\cite{comp1,comp2,comp3,comp4,comp5}. Countless experiments have validated its predictions, and quantum theory remains today one of the most successful scientific theories developed by mankind. Although very few people disagree or question the correctness of quantum formalism as a mathematical model, its foundational aspects still confound physicists even after more than 90 years since its initial formulation. Issues such as the nature of the wavefunction and its collapse (in the Copenhagen interpretation) and the state superposition, as well as entanglement, still inspire debates among physicists~\cite{debate1, solvay, speakables}. Apart from the ``standard''  Copenhagen interpretation, there are several other interpretations of the quantum formalism such as the pilot wave theories (e.g. Bohmian mechanics~\cite{bohmian1,bohmian2}), many worlds theories~\cite{Everett}, QBism~\cite{QBism}, the retrocausal  interpretations~\cite{Cramer,tsvf}, and many more. The ``apparent incompleteness'' of the wavefunction description was one of the main reasons that led to these different interpretations. Therefore, physicists suggested to augment the wavefunction with different entities, nowadays referred to as hidden variables~\cite{hvreview}. Among the different classes of hidden variables, local ~\cite{bell} and crypto-nonlocal hidden variable theories ~\cite{leggett} have been tested and ruled out experimentally, thus showing quantum mechanics to be incompatible with local realistic and even some non-local realistic theories. Alternatively, as the formalism perfectly describes the application of quantum theory to physical problems, many scientists choose to not dwell on the meaning behind the formalism, i.e. on the foundations of quantum mechanics, but rather but rather adopt the famous ``Shut up and calculate'' proverb~\cite{merminproverb}. However, some relatively new developments in closing various experimental loopholes, e.g. freedom-of-choice, fair-sampling, communication (or locality), coincidence and memory loopholes~\cite{freedomchoice,loophole,fair-sampling,Cosmic,Josephson,efficientdetection}, have led to a resurgence of interest in quantum foundations, especially in quantum entanglement, even including a global test of entanglement involving many countries, institutes and layman participants~\cite{BigBell}.

Researchers have also been striving to reach a clear understanding of what really differentiates a quantum theory from a classical theory. For instance, the concept of superposition also appears in classical wave mechanics. In Young's double-slit experiment, coherent light waves, diffracted from two slits, superpose and interfere constructively or destructively at different positions in space, resulting in bright and dark fringes at the far-field region of the slits. This experiment, easily explained by Maxwell's equations, is conceptually different when  repeated with a single photon source, or any single quantum objects. Though the probability of detecting photons in the far-field region follows the same fringes pattern, one may ask ``which slit does the photon choose to traverse through?''  One can assign some sort of unknown local physical parameters (hidden variables) which determine the path of the single photon. However, these  ``hidden variables'' are incapable of describing the experimental outcome at the single-photon regime. Even in principle, if we have some way of obtaining the which-path information, then the interference pattern is different. It is impossible to assign local hidden variables to describe the photon's whereabouts before it is actually detected on the screen. These kinds of experiments,  analyzing particles at the atomic or molecular level ~\cite{expt1,expt2,expt3,expt4,expt5,expt6,expt7,expt8}, touch the very heart of quantum foundations and, in fact, according to Feynman the two-slit experiment contains ``the only mystery'' of quantum mechanics~\cite{feynman,feynman1}. This renders quantum mechanics completely different from any classical theory, and classical electrodynamics in particular, for the above example, and also illustrates many of the questions that have been puzzling scientists since the last century. Very recently, wave superpositions among different degrees of freedom of a physical system, e.g. polarization and spatial modes of an optical beam, have been referred to as ``classical entanglement''~\cite{spreeuw,oxymoron}. We believe that the term ``classical entanglement'' is a misnomer as it can easily lead to confusion amongst non-experts and, sometimes, even experts in the field. First, classical electrodynamics can perfectly describe the physics, as well as correlations, among these different degrees of freedom of optical waves. Thus, there is no need to invoke quantum mechanics for superpositions of different degrees of freedom of light. Moreover, entanglement is the fundamental feature of quantum physics between two (or more) systems and the consequences drawn from the obtained correlations do not apply to any classical system, i.e. classical correlations cannot lead to the same conclusions as quantum entanglement. While analogies might be seen in the mathematical formulation, the possibility  of spatial separation, which is the key aspect of entanglement, does not hold for the classical counterpart.  However, it is important to point out that superposition among different degrees of freedom of a quantum object, e.g. single photon~\cite{singleparticleent,singleparticleCHSH} or neutron~\cite{neutronent}, can be used to test the contextuality of quantum mechanics, which is a \emph{rich} subject of research in itself~\cite{singleparticleCHSH}. Historically, the term entanglement (``verschr\"{a}nkt'' in German) was introduced by Schr\"odinger to describe nonlocal correlations among different quantum systems. Numerous experiments performed on multiparticle entangled states, such as the Hong-Ou-Mandel effect~\cite{hongoumandel}, the Franson interferometer~\cite{franson}, etc., have exhibited correlations that do not have any classical counterparts, thus showing entanglement to be purely a quantum effect. Here, we try to provide a comprehensive perspective on entanglement, local and crypto-nonlocal hidden variables, as well as contextual hidden variable theories. We further discuss the relatively new terminology of ``classical entanglement'' and hope to clarify its limitations. In addition to these concepts, some recent developments in understanding entanglement, e.g. improvements of nonlocal bounds and their relation to generalized uncertainty principles. We conclude with a future outlook in these areas.

\section{Popper's diffraction experiment}
\begin{figure}
	\includegraphics[width=\linewidth, angle=-0.025]{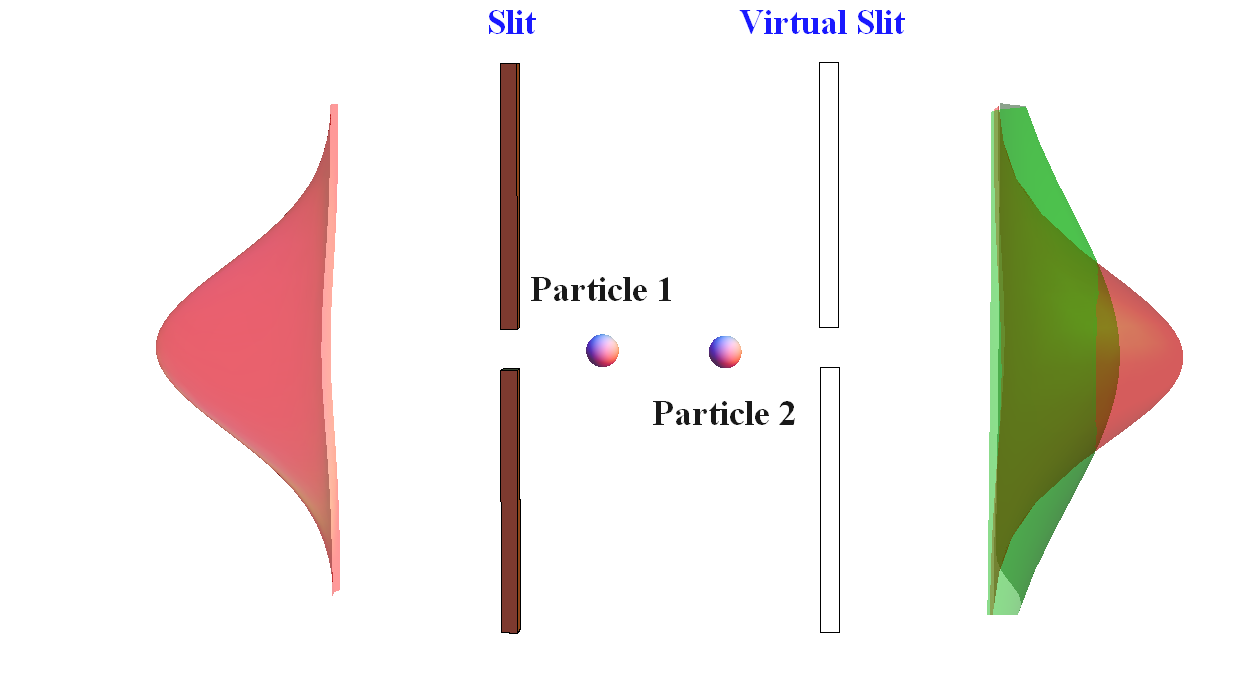}
	\caption{Popper's ghost diffraction setup. Two entangled particles which we label as 1 and 2 travel in opposite directions from a source S. One of the particles, say particle 1, passes through a slit. We then look at the effect of it, i.e. changing the uncertainties in the position of particle 1, on the momentum spread of particle 2. The green curve is the original particle wavefunction, and the red curve is the second particle's wavefunction after particle 1 passes through the slit. While  particle 2 is more localized, the momentum spread remains constant, thereby saturating the uncertainty relation, but not violating it.}
	\label{PopperSetup}
\end{figure}

In 1934, Karl Popper proposed a thought experiment~\cite{karlpopper} with entangled particles aimed at analyzing the correctness of the Copenhagen interpretation of quantum mechanics. In his experiment (Fig. \ref{PopperSetup}), he considered a pair of  particles entangled in position and transverse momentum that are traveling in opposite directions towards the two slits. When one of the particles  passes through a slit, then by virtue of entanglement we also acquire  position information of the second particle. Popper then analyzed the two possible ways the  measurement result of the momentum of the second particle could unfold. The first one, which he argued, is  according to his understanding of the Heisenberg uncertainty principle, is that  position measurement of first particle should cause a large spread in the momentum of the second particle. Heisenberg uncertainty principle for position and momentum  of a single particle states, 
\begin{equation}	
\Delta x\Delta p\geq\frac{\hbar}{2},
\end{equation}
where $\Delta x$ and $\Delta p$ are the uncertainties in the position and momentum, respectively.  We see that, according to the Heisenberg uncertainty relation, when the position is known precisely there is a large spread in the momentum.
However, this violates the principle of causality as, due to the narrowing or widening of the slit for the first particle, we would instantaneously affect the momentum spread of the second particle. Therefore, according to Popper, we are presented with the choice between relativistic causality and the Copenhagen interpretation. Popper suggested that the momentum spread would not change in an experiment and came to the conclusion that the Copenhagen interpretation must be inadequate \cite{Popperbook}.

Looking at this experiment more closely, when considering the uncertainty relation for the second particle, we should use the uncertainties in position and momentum that are conditioned upon the measurement result of the first particle. The uncertainty relation reads,
\begin{equation}
\Delta(x_2|x_1) \Delta(p_2|x_1)\geq \frac{\hbar}{2},
\end{equation} 
where $\Delta(x_2|x_1)$ and $\Delta(p_2|x_1)$ are the uncertainties  in the position and the momentum of the second particle, respectively conditioned upon the outcome of the position measurement of the first particle. A recent experiment performed using entangled photons generated by spontaneous parametric down conversion (SPDC)~\cite{popperexperiment,miles:2018}, has shown that, while there is no spreading of the wavefunction, it is still consistent with the standard quantum formalism  and the conditioned uncertainty principle.
{While there is less uncertainty in position, the momentum uncertainty remains the same and the product saturates the uncertainty relation but does not violate it.} Several other experiments analyzing the different aspects of Popper's experiment have also vindicated the correctness of standard quantum formalism~\cite{Shih,p:18,popperexperiment,miles:2018}.
Popper's experiment first showed that entangled states raise profound questions about the nature of quantum mechanics, and indeed we will see in the subsequent sections how entanglement plays a fundamental role in quantum foundations.

\section*{EPR and Local realism}
\     In 1935, Albert Einstein, Boris Podolsky and Nathan Rosen (EPR) first considered the now well-known EPR pair of particles in their highly influential paper~\cite{EPR}, although without using the term ``entanglement''. Analyzing these entangled states, they questioned the completeness of quantum mechanics. In line with the EPR argument, a theory is complete only if it has a physical quantity corresponding to each element of reality. As defined by EPR, a physical quantity is real if its value can be predicted with certainty irrespective of and before any measurement. For example, in classical mechanics, the set of position and momentum (or velocity) is sufficient to  assign definite values to any other dynamical physical quantities such as kinetic energy, angular velocity, etc. In this sense, the description provided by classical mechanics can be considered a complete characterization of reality for the particle. EPR questioned whether  the wavefunction (or the state vector) in quantum mechanics is a complete description of physical reality. In quantum mechanics, two physical quantities represented by non-commuting observables cannot be measured simultaneously with arbitrary precision. Whenever we measure one observable, we influence the state in such a way that the measurement outcomes for the other observable is disturbed. Therefore, simultaneous ``realities'', at least as per the EPR criteria, do not exist for non-commuting observables. Thus, EPR argued that either $a)$ quantum theory is incomplete because it cannot simultaneously describe the reality of both of these observables, or $b)$ there is no simultaneous reality of two non-commuting observables. By using an example of an entangled system with two particles they claimed that there exist two different simultaneous realities for a physical system according to quantum theory. Thus, EPR concluded that quantum theory must be incomplete \textit{in its current form}. Note that EPR excluded the possibility of measurements between space-like separated events affecting each other instantaneously, or in Einstein's words ``spooky action at a distance'' \cite{spooky}, as in their opinion this would contradict special relativity. We will come back to this important issue of nonlocality later in this section.

Although EPR phrased their argument in terms of position and momentum correlation (just like Popper), it is more useful in the context of this manuscript to use the simpler example of particles entangled in the spin degree of freedom introduced by Bohm ~\cite{bohm,aharonovbohm}. Consider an anti-correlated spin state of two particles \emph{A} and \emph{B}, e.g. generated via spontaneous decay,
\begin{equation}
\ket{\psi}= \frac{1}{\sqrt{2}}\left(\ket{\uparrow}_{A}\ket{\downarrow}_{B}-\ket{\downarrow}_{A}\ket{\uparrow}_{B}
\right),
\end{equation}
where $\ket{\uparrow}$ and $\ket{\downarrow}$ are \emph{spin up} and \emph{spin down} states in the  $z$ direction respectively. Note that this formalism can be extended to any other two-dimensional vector spaces, e.g. photonic polarisation or path.

Let us assume that the two particles \emph{A} and \emph{B} are spatially separated, such that any local \emph{physical} interaction between them is circumvented. One can perform a measurement on the spin state of particle \emph{A} in two different bases, say eigenstates of the $\hat{\sigma}_z$ and $\hat{\sigma}_x$ operators,  similar to the position and momentum basis in the original EPR argument. Upon performing these measurements, two scenarios will arise for the spin state of particle \emph{B}:
\begin{itemize}
	\item [$\hat{\sigma}_z$:] Depending upon the outcome of the particle \emph{A} spin-state measurement, the spin state of particle \emph{B} is either $\ket{\uparrow}_B$ or $\ket{\downarrow}_B$ -- it is always opposite to the particle \emph{A} spin-state, since the two particles are anti-correlated in the spin degree of freedom. Upon finding the particle  \emph{A} in, say $\ket{\uparrow}_A$, according to EPR, since the first particle cannot affect the second, the state of the second particle should be $\ket{\downarrow}_B$ and the spin in $z$ direction has a value of $-{\hbar}/{2}$.
	
	\item [$\hat{\sigma}_x$:] Now we perform  the measurement in the eigenbasis of the $\hat{\sigma}_x$ operator, i.e. $|\pm\rangle = \frac{1}{\sqrt{2}}\left( \ket{\uparrow}\pm\ket{\downarrow}\right)$. The original state written in this basis is,
	\begin{equation}
	\ket{\psi} = \frac{1}{\sqrt{2}}\left( \ket{+}_{A}\ket{-}_{B}-\ket{-}_{A}\ket{+}_{B}
	\right).
\end{equation}
Note that this state has the same form as in the $\ket{\uparrow},\ket{\downarrow}$ basis. In fact, the original state has the same form in any orthogonal basis, and we call such states rotationally invariant. Let us assume that we observe particle \emph{A} in the $\ket{+}$ state. Again we do not disturb the second particle and thus we conclude particle \emph{B} to be in the state, $\ket{-}=\frac{1}{\sqrt{2}}\left(\ket{\uparrow} - \ket{\downarrow} \right)$. This is an eigenstate of $\hat{\sigma}_x$ with spin $-{\hbar}/{2}$. 
\end{itemize}
Without in any way disturbing or interacting with the second particle, we have obtained two simultaneous spin values and states for $\hat{\sigma}_{z}$ and $\hat{\sigma}_{x}$, e.g. $\ket{\psi}_B=\ket{\downarrow}$ and $\ket{\psi'}_B=\ket{-}=\frac{1}{\sqrt{2}}\left(\ket{\uparrow} - \ket{\downarrow} \right)\neq\ket{\psi}_B$. Therefore, EPR claim that it is possible to assign the spin values for the two non-commuting operators. This means that the state vector description of quantum mechanics must be an incomplete description of reality.

Nowadays, we understand that there are several problems with the EPR reasoning. One is that a single particle state vector is not an accurate description of the single particle when it is in an entangled state. Quantum mechanics resolves the ambiguity in spin values and the state representation of a particle in an entangled state by using density matrices, which provide a more complete way of representing mixed states. In the density matrix formalism, the joint state of the two particles is represented by the density matrix,
\begin{align*}
\hat{\rho}_{AB} &= \ket{\psi}\langle \psi| \\
& =\frac{1}{2}\left(\ket{\uparrow}_A\ket{\downarrow}_B-\ket{\downarrow}_A\ket{\uparrow}_B\right)\,\left(\bra{\uparrow}_A\bra{\downarrow}_B-\bra{\downarrow}_A\bra{\uparrow}_B\right).
\end{align*}
If we consider only  particle \emph{B}, its density matrix is,
\begin{align}
\hat{\rho}_{B}&= \text{Tr}_\text{A}\left(\ket{\psi}\bra{\psi}\right) \nonumber \\
& = \frac{1}{2}(\ket{\uparrow}_B\!\bra{\uparrow}+\ket{\downarrow}_B\!\bra{\downarrow})=\frac{1}{2}\,\hat{\mathbb{I}},
\end{align}
where $\hat{\mathbb{I}}$ is the identity operator.\newline Similarly if we decide to measure in the $\ket{+}, \ket{-}$ basis,
\begin{align}
\hat{\rho}_{AB}&= \ket{\psi}\bra{\psi} \nonumber\\
& =\frac{1}{2}\left(\ket{+}_A\ket{-}_B-\ket{-}_A\ket{+}_B
\right)\left(\bra{+}_A\bra{-}_B-\bra{-}_A\bra{+}_B
\right). \nonumber\\
\hat{\rho}'_{B} &=  \text{Tr}_\text{A}\left( \ket{\psi} \bra{\psi} \right) \nonumber\\
& =\frac{1}{2}(\ket{+}_{B}\!\bra{+}+\ket{-}_{B}\!\bra{-} ), \nonumber\\
& =\frac{1}{2}\hat{\mathbb{I}}.
\end{align}
Hence, we observe that density matrices resolve the ambiguity in the state representation for photon \emph{B}. Photon \emph{B} possesses a unique density matrix, $\hat{\rho}'_{B}=\hat{\rho}_{B}=\hat{\mathbb{I}}/2$  what we refer to as maximally mixed state, independent of performing a measurement on photon \emph{A} spin state.\newline The state of photon \emph{B} if considered independently of \emph{A} is a mixed state. On the contrary, as described before, a measurement conditioned on the outcome of photon A leads to a perfectly predictable outcome for photon \emph{B}, i.e. perfect correlation in any bases.

There is another fundamental issue of nonlocality pertaining to entangled states: the idea that measurements performed in spatially separated locations can affect each other. EPR assumed that nature is local and believed that it would violate the principle of causality if experiments performed in one location could affect experiments in far away places.  As we will discuss, Bell later showed that a local-realistic description of entangled states is inconsistent with quantum mechanics, effectively ruling out the local hidden variables description of entangled states~\cite{bell}.

\section*{Schr\"odinger's cat state and ``entanglement''}
Erwin Schr\"odinger, inspired by the EPR paper, introduced the term ``entanglement'' for the first time in his 1935 paper~\cite{scat}. He used the phrase ``entanglement of our knowledge of the bodies'' to refer to the joint states where the state of one system is intrinsically linked with the state of another system. Additionally, he also provided a thought experiment involving a macroscopic (classical) object, namely a cat, to illustrate the nature of quantum superpositions in entangled states. 
In his experiment, a cat is placed inside a steel chamber with a small amount of radioactive material. The radioactive material is coupled with a Geiger counter and a vial of poisonous hydro cyanic acid. With a finite probability, one atom of the radioactive material may decay in the course of next few hours, which then registers a click in the Geiger counter. A mechanical apparatus is arranged such that once the counter clicks, it smashes the bottle of hydro cyanic acid releasing the poisonous gas, which kills the cat.

Initially, when no atom has decayed and the cat is alive, the states of the atom and the cat are:
\begin{align}
	&\ket{\psi}_{\text{atom}} = \ket{\text{no\mbox{-}decay}}  ,
    &\ket{\psi}_{\text{cat}} = \ket{\text{alive}}.	
\end{align}

 After some time, it is now impossible to say whether any atom has decayed or not. Depending upon the state of the radioactive atom, it is impossible in turn to ascertain whether the cat is alive or dead. So, until an external observer checks whether the atom has decayed or not, the cat is in a \emph{weird} state of being \emph{dead} and \emph{alive} at the same time. The joint state of the atom and the cat is,
\begin{eqnarray}
\ket{\psi} &=& \ket{\text{Atom and Cat}}\cr
&=&\frac{1}{\sqrt{2}}\left(|\text{decay}\rangle|\text{dead}\rangle + |\text{no-decay}\rangle|\text{alive}\rangle\right).
\end{eqnarray}
In the words of Schr\"odinger, our knowledge of the two bodies becomes entangled. He further elaborated the definition of entanglement by adding that, even if the two bodies are taken very far from each other, ``knowledge of the two systems cannot be separated into  the logical sum of knowledge about two bodies''. It is not possible to express the state of the cat and the atom independently of each other. Until an external observer opens the door of the chamber, thereby performing the measurement, the state remains in the superposition.
 
 Schr\"odinger initially proposed this thought experiment linking the microscopic world to the macroscopic, classical world to question the Copenhagen interpretation claiming such ``blurred'' states where the cat is in some kind of superposition of being alive and dead can only be observed in the microscopic world. It raises the question when the macroscopic objects stop being in a superposition and transform into either one or the other of the alternatives, and if this transition needs to happen at all. This is still an ongoing debate, and there are numerous experiments pushing the limits of ``macroscopic'' superpositions~\cite{macro1,macro2,macro3,macro4}. More importantly, this experiment illustrates the nature of quantum superposition in the context of entangled states.
 
 Wigner~\cite{wignersfriend} proposed an extension of this experiment, with two observers. One observer, say Wigner, stays outside the chamber and another, say Wigner's friend, is positioned inside the chamber. Wigner's friend, by virtue of being inside the chamber can observe a definite outcome, i.e. whether the cat is dead or alive. On the other hand, Wigner has no way of knowing the outcome until his friend mentions it to him. For Wigner, the joint state of the whole system is,
 \begin{eqnarray}
 \ket{\psi} &= \frac{1}{\sqrt{2}}\big[\ket{\text{no-decay}}\ket{\text{alive}}\ket{\text{friend sees alive cat}} \nonumber \\
 &+\ket{\text{decay}}\ket{\text{dead}}\ket{\text{friend sees dead cat}} 
 \big].
 \end{eqnarray}
 The paradox occurs when we ask ``when did the cat stop being in a superposition state?'' For Wigner's friend, it occurs whenever he decides to check if the cat is alive. Wigner, however, sees the state in superposition until his friend tells him the outcome. The two observers will not agree about the time when the cat will be in a definite state. Until today, similar arguments are discussed with novel twists \cite{Vedral1,Vedral2,hw:2019,wignerbruckner,wignerrenner}, which show that the perplexing nature of the entanglement between a ``macroscopic''(or classical) and a ``microscopic'' (or quantum) object is still worth a discussion. 
\section{Hidden Variables}
\begin{figure}
	\includegraphics[width=\linewidth]{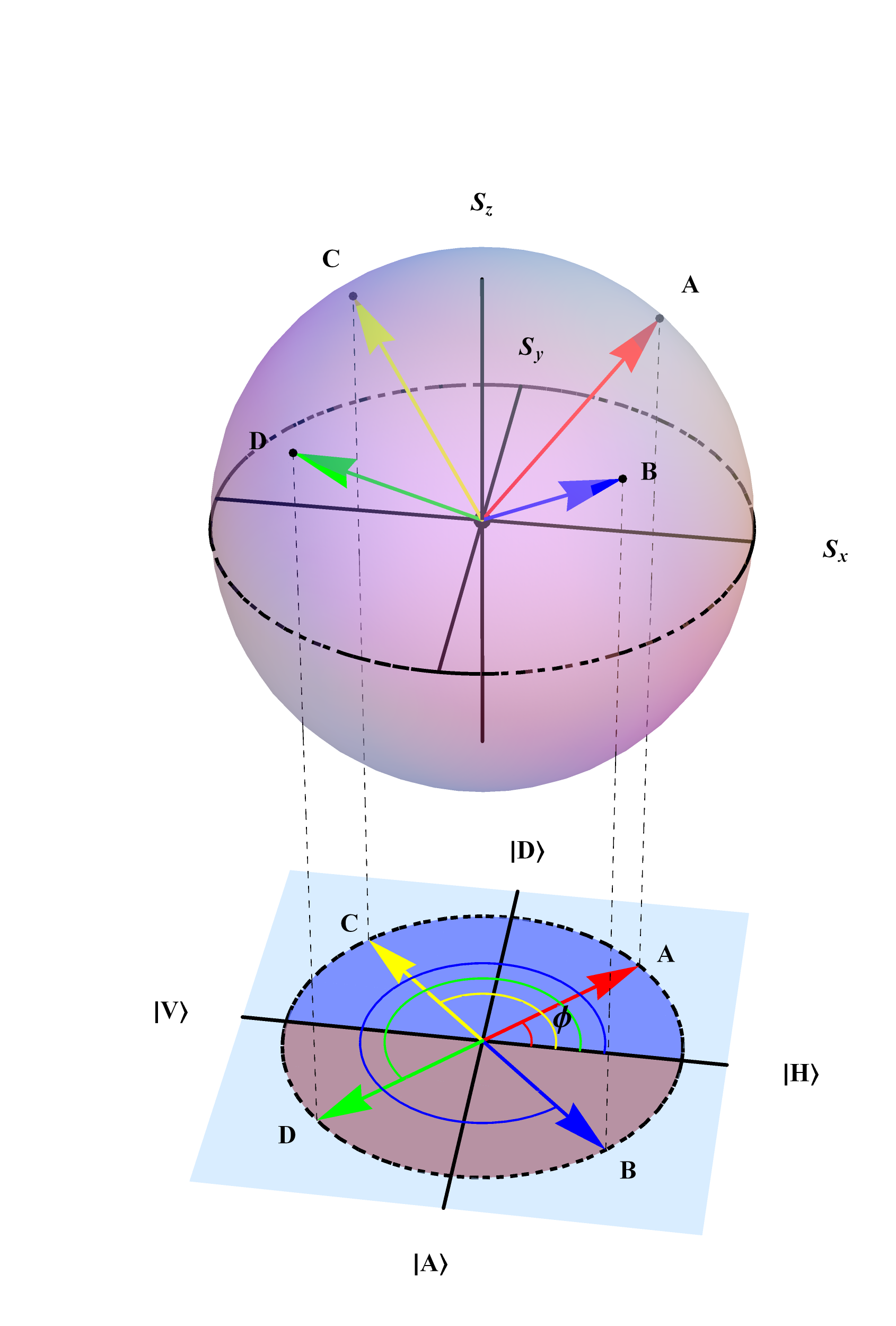}
	\caption{Four polarization-vectors A, B, C  and D are represented in Poincar\'{e} sphere. The angles made by each vector with the y-axis, which determines which half of the x-y plane each vector lies, could the hidden variables for polarization components in the other two directions (x and y). Here the vectors A and C with projections in the region shaded by blue have positive polarization in the x direction, while vectors B and D with projections in the region shaded by red have negative polarization in x direction.}
	\label{3D-Polarization}
\end{figure}

In the spirit of EPR, who argued that quantum mechanics is incomplete, hidden variable theories attempt to supplement quantum mechanics by adding some extra parameters. These  parameters, or the so called ``hidden variables'', are assumed to exist beyond the standard formalism of quantum mechanics and are supposed to resolve the probabilistic nature of quantum experiments. In other words, the hidden variables ensure that in principle there can be a deterministic description of all observables, which might just be unknown to us, thereby restoring the realistic (in the above mentioned sense) description of nature. To look at a simple example of how such hidden variables work, consider, for instance, a photon's polarization. We could represent the polarization of a photon by a three dimensional vector in a sphere called the Poincar\'{e} sphere, with each dimension representing the corresponding polarization component. Let us look at four such photons whose polarization vectors make the same angle, $\theta<\frac{\pi}{2}$ as shown in Fig. \ref{3D-Polarization}, with  the $z$-axis, implying that they all have a positive $z$ component. However, along the $x$-axis and $y$-axis they can have different polarization. In general, there can be infinitely many such vectors with the same $z$ component that form a cone around the $z$-axis. Standard quantum mechanics postulates that for the particles measured to have positive polarization in the $z$-direction, the polarization measurement along say the $y$ axis can give rise, probabilistically, to either the positive, (meaning diagonal polarization $\ket{D}$) or negative (meaning antidiagonal polarization $\ket{A}$). However, in the classical picture we considered, one can assign a hidden variable $\lambda:=\lambda(\phi)$, such that,
\begin{equation}
  \lambda(\phi) =
  \begin{cases}
                                   +1 & \text{if $0\leq\phi\leq\pi$} \\
                                   -1 & \text{if $\pi\leq\phi\leq 2\pi$}
  \end{cases}. 
\end{equation}
The variable $\lambda$ would then accurately describe the polarization in the $y$ direction.
For the vectors with $\lambda=+1$, i.e. whose projection vector in the $x-y$ plane lies on the positive half-plane (the region shaded blue in Fig. \ref{3D-Polarization}),the polarization in the $y$ direction is positive.  For the vectors with $\lambda=-1$, i.e. whose projection vector in the $x-y$ plane  lies on the negative half-plane (the region shaded red in Fig. \ref{3D-Polarization}) the polarization in the $y$ direction is negative. Here, $\lambda$ serves an example of a hidden variable for the polarization along $y$-direction.

Different variations of the hidden variables exist depending on their properties and the dependence of these variables on measurement settings in a quantum experiment. We present below some classes of hidden variables theories, a few of which have already been proven to be incompatible with the predictions of quantum theory.

\subsection*{Local hidden variables and Bell's proof}
As the name implies, a local hidden variables theory assumes that the hidden variables for a particular quantum system are local and  unaffected by experiments performed at any other spatially separated locations. To define it more formally, consider an EPR-like setup with two space-like separated and strongly correlated particles A and B. Again let $\lambda$  be the hidden variable, and  \textbf{A} and \textbf{B} the spin values of A and B measured along directions \textbf{a} and \textbf{b}, respectively. Then according to a local hidden variables theory, the measurement result \textbf{A} depends only upon the  measurement setting \textbf{a} and the hidden variable $\lambda$. It is independent of B's measurement setting \textbf{b}, and similarly for B, 
\begin{align}
   \label{ABlhv}
   \textbf{A} &=\textbf{A}(\lambda,\textbf{a}), \\ \nonumber
   \textbf{B} &=\textbf{B}(\lambda,\textbf{b}).  
\end{align}
We also assume that the hidden variable has probability distribution $\rho(\lambda)$. Apart from these, we do not make any assumptions about the nature of the hidden variable or its probability distribution. Note that by such local hidden variables, the only assumption made are locality and realism; hence, at this stage, no knowledge of quantum mechanics has to be known. Then, in this hidden variable formalism, the two particle correlation, or the expectation value of the product of the two observations of \textbf{A} and \textbf{B}, is then given by,
\begin{equation}\label{lhv}
P(\textbf{a},\textbf{b})= \sum_{\lambda} \rho(\lambda)\,\textbf{A}(\textbf{a},\lambda)\,\textbf{B}(\textbf{b},\lambda).
\end{equation}   

In his seminal paper~\cite{bell}, John Bell proved that any theoretical prediction for  measurement outcomes fulfilling the ideas of locality and realism is upper bounded for a given set of measurements (the so-called Bell inequality). He also showed that quantum mechanics allows for the possibility to exceed this bound proving that quantum correlations cannot be obtained from any local realistic hidden variable theories with the form described by Eq.~(\ref{lhv}).
Since the time of his paper, numerous experiments have been performed that attest to the correctness of quantum mechanics and falsified the assumption of local hidden variables \cite{Brunner,fair-sampling,Cosmic,Josephson,efficientdetection,BigBell}, most recently even loophole-free \cite{loophole,PhotonPRL1,PhotonPRL2}. Contrary to the EPR assumption, nature does seem to allow the measurement of one particle to affect the ``reality'' of the other.  

\begin{figure}
	\includegraphics[width=\linewidth]{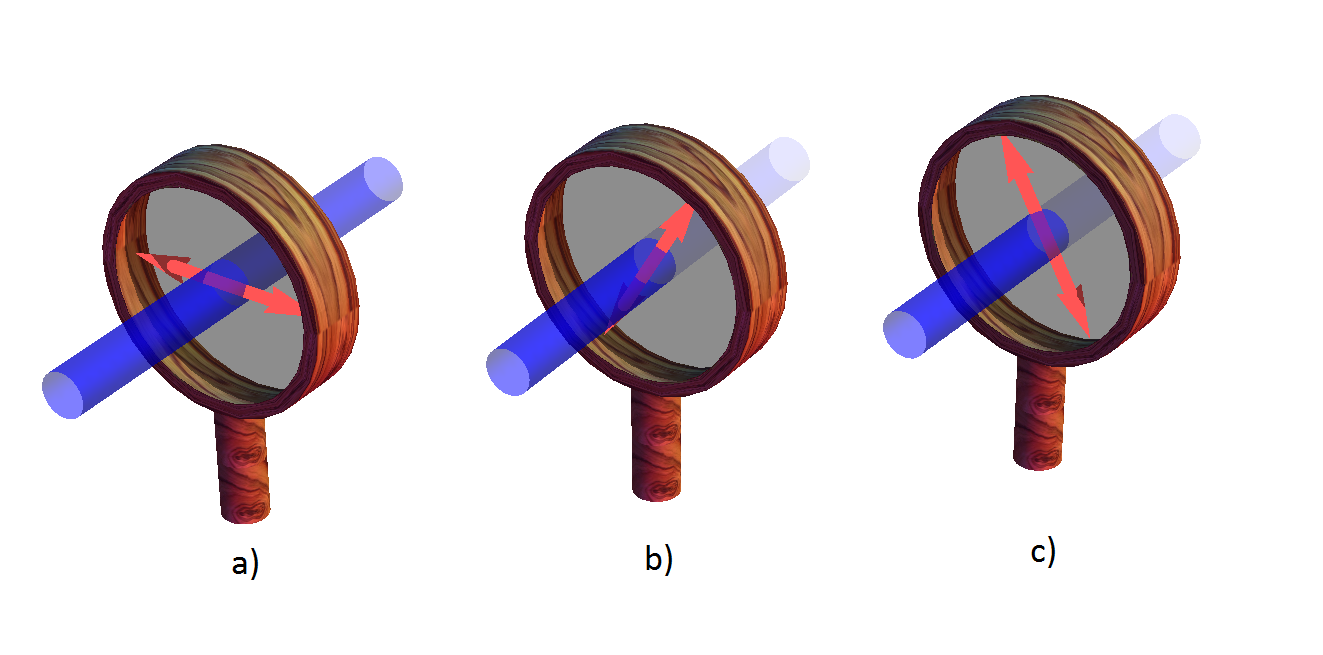}
	\caption{Three different directions for the polarization measurements: \textbf{(a)} $0$, \textbf{(b)}$-2\pi/3$, and \textbf{(c)}$+2\pi/3$.}
	\label{basis}
\end{figure}

We present here a simpler proof of incompatibility of local hidden variables with quantum mechanics. Consider a source that prepares a pair of photons with perfectly correlated polarization in any direction, i.e. if one is horizontally polarized then the other is also horizontally polarized and so on. In our experiment, we are able to measure the polarization of each of the two photons, at angles $0$, $+2\pi/3$, and $-2\pi/3$ using three different polarizer settings (Fig \ref{basis}).

Assuming that the polarization results in these three directions are pre-defined by local hidden variables before the measurement, we tabulate all the eight possible combinations of the hidden variables for each photon (Table \ref{table1}).

\begin{table}[ht]
    \centering
	\caption{\textbf{Single photon polarization values in three different directions in local hidden variable theory.} \checkmark~means the photon will pass through the polarizer set in that direction and \xmark~means that it will be  blocked by the polarizer.}
	\begin{tabular}{@{} |p{1.8cm}|p{2.0cm}|p{2.2cm}|p{2.2cm}| @{}} \hline
		\textbf{Outcomes} & $\mathbf{a_1/b_1} (0)$ & $\mathbf{a_2/b_2} (+2\pi/3)$ 
		& $\mathbf{a_3/b_3} (-2\pi/3)$ \\ \hline
		1 & \cmark & \cmark & \cmark \\ \hline
		2 & \cmark & \cmark & \xmark \\ \hline
		3 & \cmark & \xmark & \cmark \\ \hline
		4 & \cmark & \xmark & \xmark \\ \hline
		5 & \xmark & \cmark & \cmark \\ \hline
		6 & \xmark & \cmark & \xmark \\ \hline
		7 & \xmark & \xmark & \cmark \\ \hline
		8 & \xmark & \xmark & \xmark \\ \hline
	\end{tabular}
\label{table1}
\end{table}

We then look at the results of Alice's and Bob's measurements in two different directions  and  note whether they get the same results or not (Table \ref{table2}).
\begin{table}[ht]
	\centering
	\caption{\textbf{Two photon polarization measurement results.} $\mathbf{a_1\,b_2}$ refers to the polarizer for the first photon is set at $0$ and that for second photon is set at ${3\pi}/{4}$, and so on. \checkmark~means that both photons have same outcomes for the directions specified, i.e. either both photons pass through the specified polarizers or both are blocked. \xmark~means that they have different outcomes.}
	\begin{tabular}{@{} |p{1.8cm}|p{1.5cm}|p{1.5cm}|p{1.5cm}|p{1.5cm}| @{}} \hline
		\textbf{Outcomes} & $\mathbf{a_1,b_2}$ &
		$\mathbf{a_2,b_3}$ &
		$\mathbf{a_3,b_1}$ &
		$\mathcal{P}_\text{same}$ \\ \hline
		1 & \cmark & \cmark & \cmark & $1$ \\ \hline
		2 & \cmark & \xmark & \xmark & $1/3$ \\ \hline
		3 & \xmark & \xmark & \cmark & $1/3$ \\ \hline
		4 & \xmark & \cmark & \xmark & $1/3$ \\ \hline
		5 & \xmark & \cmark & \xmark & $1/3$ \\ \hline
		6 & \xmark & \xmark & \cmark & $1/3$ \\ \hline
		7 & \cmark & \xmark & \xmark & $1/3$\\ \hline
		8 & \cmark & \cmark & \cmark & $1$\\ \hline
	\end{tabular}
	\label{table2}
\end{table}
From the table, we can deduce that the probability of seeing the same result is always at least 1/3. To see why this is true, let us assume we are dealing with the pair of photons represented by the second row of the table. Now if we randomly perform the measurement then, one third of the time we would be measuring in the $\mathbf{a_1b_2}$ basis and we would see the states to be the same. In the remaining two-thirds, we would see the states to be different. Looking at  rows 2-7 we see that each of them has the same probability, 1/3, for observing the same outcome. For the cases in rows $1$ and $8$, we get the same results every time. Hence if we assume the existence of local hidden variables, then we should see the same results at least one third of the time. One subtle point to note here is that
we have not made any assumptions about how often each of the eight possible combinations occurs in nature. Each of the eight possible combinations of the hidden variables could occur with any probability and still the result would be the same. Therefore, according to a local hidden variable (LHV) theory, the probability for observing the same outcome is bounded as,
\begin{equation}
\mathcal{P}(\text{LHV, same}) \geq \frac{1}{3}.
\end{equation}
Now we turn to quantum mechanics and what it tells us about such states. 
For a source that produces photons with perfectly correlated polarizations in any direction, the initial two-photon state is an entangled state which is written as,
\begin{equation}
\ket{\psi}=\frac{1}{\sqrt{2}}\big(\ket{H}_A\ket{H}_B + \ket{V}_A\ket{V}_B \big),
\end{equation}
where $\ket{H}$ and $\ket{V}$ refer to the horizontal and vertical polarization respectively.

 Let us denote the three measurement directions (or polarization bases) for A by \textbf{a\textsubscript{i}}, and for B by \textbf{b\textsubscript{i}}. Written in terms of $\ket{H}$ and $\ket{V}$,
\begin{equation}
\begin{aligned}
&\ket{\mathbf{a_1}}=\ket{H}, \\
&\ket{\mathbf{a_2}}=-\frac{1}{2}\ket{H}+\frac{\sqrt{3}}{2} \ket{V}, \\
&\ket{\mathbf{a_3}}=-\frac{\sqrt{3}}{2} \ket{H}-\frac{1}{2} \ket{V},
\end{aligned}
\end{equation}
and similarly for B.

Then the probability of observing the same polarization in the same basis is given by the cosine of the angle 
between the two states, which is $2\pi/3$ in our case. Hence, according to Quantum Mechanics(QM),
\begin{equation}
\mathcal{P}(Q\text{M, same}) = \cos^{2}(2\pi/3) = 0.125.
\end{equation}
Thus, we observe that the probability obtained from the assumption of local hidden variables contradicts the probability derived by quantum mechanics. Experimentally, the results have always vindicated the predictions of quantum mechanics, thus favoring quantum mechanics over local hidden variables theories.
Alongside with many Bell inequalities~\cite{Brunner} that were proposed to rule out local hidden variables models, it is worth mentioning the existence of proofs which do not invoke inequalities~\cite{GHZ1,GHZ2,HardyNonlocality}.

\subsection*{Nonlocal hidden variables theories}
Until now we restricted the hidden variable model to be local, which is an assumption well-justified by another major physical theory known today, i.e. general relativity. However, as we have seen above, local hidden variable models cannot explain the correlations in entangled states; thus, it might be a natural thing to ask if the locality assumption is too strong and can be lifted to find an agreement with quantum mechanics. Leggett laid out a more general nonlocal hidden variable theory~\cite{leggett}, which assumes that:
\begin{itemize}
	\item[1.] Each pair of photons in an EPR-like setup is characterized by a hidden variable $\lambda$.
	\item[2.] The distribution of the hidden variable $\lambda$, $\rho(\lambda)$, is independent of the measurement settings \textbf{a}, and \textbf{b} and the results of the measurements \textbf{A}, and \textbf{B} of either of the particles. 	
	\item[3.] The results of each measurements \textbf{A} and \textbf{B} depend upon the hidden variable $\lambda$, as well as both of the measurement settings \textbf{a} and \textbf{b}, and the results of the measurement performed on the other particle. I.e., 
\begin{align}
   \textbf{A} &=\textbf{A}(\textbf{a},\textbf{b},\lambda,\textbf{B}), \\
   \textbf{B} &=\textbf{B}(\textbf{a},\textbf{b},\lambda,\textbf{A}).  
   \end{align}
\end{itemize}

Previously, in the local hidden variable theory, Eqs.~(\ref{ABlhv}), we saw that the outcomes \textbf{A} and \textbf{B} depend only on the hidden variable $\lambda$ and the respective measurement settings. In a local theory, the measurement settings of A and its results cannot affect the outcomes of B and vice versa, as it assumes that space-like separated events cannot influence each other. In contrast, a nonlocal hidden variable theory assumes nature is non-local, and, consequently, the outcomes for A and B are dependent not only upon their respective measurement settings, but also upon the setting of the other party and their outcomes.
	In such a theory, the expectation value of the product of the two outcomes is then given by,
\begin{equation}
	P(\textbf{a},\textbf{b})= \sum_{\lambda} \rho(\lambda)\,\textbf{A}(\textbf{a},\textbf{b},\lambda,\textbf{B})\,\textbf{B}(\textbf{a},\textbf{b},\lambda,\textbf{A}),
\end{equation}

or for the case of a continuous hidden variable  $\lambda$,
\begin{equation}
	P(\textbf{a},\textbf{b})= \int_{\lambda}d\lambda\,\rho(\lambda)\,\textbf{A}(\textbf{a},\textbf{b},\lambda,\textbf{B})\, \textbf{B}(\textbf{a},\textbf{b},\lambda,\textbf{A}).
\end{equation}   

Such nonlocal hidden variable models can describe any correlations possible-they can both give rise to the quantum mechanical predictions, for e.g. Bohmian mechanics~\cite{bohmian1,bohmian2}, or in some cases even exceed the correlations given by quantum mechanics. Although some of them have to be considered non-physical, they are interesting lines of thought themselves. \cite{nonlocalboxes1,nonlocalboxes2}
To enable a test of a subclass of such Non Local Hidden Variables (NLHV) models called crypto-nonlocal theories, Leggett added another condition, namely:

\begin{itemize}
	\item[4.] The outcomes \textbf{A} and \textbf{B} each depend upon the measurement setting of the other but are independent of the outcome, 
	\begin{eqnarray*}
	\textbf{A}(\textbf{a},\textbf{b},\lambda,\textbf{B}) =&  \textbf{A}(\textbf{a},\textbf{b},\lambda),\\ \textbf{B}(\textbf{a},\textbf{b},\lambda,\textbf{A}) =& \textbf{B}(\textbf{a},\textbf{b},\lambda).	
	\end{eqnarray*}
\end{itemize}

Let us look at a nonlocal hidden variable model \cite{nonlocalrealism1} that satisfies this condition.
We will see how this model successfully recreates quantum correlations for photons when the polarization measurement vectors are confined to a certain plane in the Poincar\'{e} sphere. As a consequence, for measurements performed in that plane, the model can even account for the violation of the CHSH inequality~\cite{CHSHineq}, an inequality which is satisfied by local hidden variable theories. However once we start performing measurements in a different plane, this model fails to recreate the quantum correlations. We will then look at the Leggett inequality that bounds correlations obtained from this type of nonlocal hidden variable model, which nevertheless is violated by quantum correlations. 

Our two parties, Alice and Bob each share a pair of photons A and B, with the initial polarization vectors \textbf{u} and \textbf{v}, respectively. We denote the polarization measurement vectors by \textbf{a} and \textbf{b}, respectively, for A and B. The measurement outcomes \textbf{A} and \textbf{B} both are binary valued ($\pm 1$) variables and as required for a crypto-nonlocal theory, do not depend upon each other. The hidden variable model predicts the measurement outcome for A as follows,
\[
  \textbf{A} =
  \begin{cases}
                                   +1 & \text{if $0\leq\lambda\leq\lambda_{A}$} \\
                                   -1 & \text{if $\lambda_A\leq\lambda\leq 1$}
  \end{cases},
\]
where $\lambda_A$ depends upon A's initial polarization and the measurement setting as,  
\begin{equation}
    \lambda_A = \frac{1}{2}(1+\mathbf{u}\cdot\mathbf{a}).
\end{equation}
Similarly for B,
\[
  \textbf{B} =
  \begin{cases}
                                   +1 & \text{if $x_1 \leq\lambda\leq x_2$} \\
                                   -1 & \text{if $0\leq\lambda\leq x_1 \;        \& \; x_2\leq\lambda\leq 1 .$} 
  \end{cases}
\]
Now if we choose the parameters $x_1$ and $x_2$ such that,
\begin{eqnarray}
    \label{x1}
    x_1 = \frac{1}{4}(1+\mathbf{u}\cdot\mathbf{a}-\mathbf{v}\cdot\mathbf{b} + \mathbf{a}\cdot\mathbf{b}), & \\
    \label{x2}
    x_2 = \frac{1}{4}(3+\mathbf{u}\cdot\mathbf{a}+\mathbf{v}\cdot\mathbf{b} + \mathbf{a}\cdot\mathbf{b})
\end{eqnarray}
then the model reproduces Malus' law,
\begin{eqnarray*}
    \langle A \rangle&=& \int_{0}^{\lambda_A}d\lambda - \int_{\lambda_A}^{1}d\lambda = \mathbf{u}\cdot\mathbf{a}, \\
    \langle B \rangle &=& \int_{x_1}^{x_2}d\lambda - \int_{0}^{x_1}d\lambda - \int_{x_2}^{1}d\lambda  = \mathbf{u}\cdot\mathbf{b}.
\end{eqnarray*}

Also, the expectation value of the product of \textbf{A} and \textbf{B} is given by,

\begin{equation}
   \langle AB \rangle = -\int_0^{x_1}d\lambda +\int_{x_1}^{\lambda_A}d\lambda - \int_{\lambda_A}^{x_2}d\lambda + \int_{x_2}^{1}d\lambda= -\mathbf{a}\cdot\mathbf{b},
\end{equation}
which is the same expression as obtained from quantum mechanics. However, the problem with this model is that it is inconsistent for some measurement directions. To see how, we first note that the variables $x_1$ and $x_2$ also have to satisfy the condition,
\begin{equation}
    0\leq x_1,\;x_2 \leq 1.
\end{equation}
Substituting the expressions for $x_1$ and $x_2$, Eqs.~(\ref{x1}, \ref{x2}), in the above equation, leads to the following inequality,
\begin{equation}
    \abs{\mathbf{a}\cdot\mathbf{b} \pm \mathbf{u}\cdot\mathbf{a}}\leq 1 \mp \mathbf{v}\cdot\mathbf{b}.
\label{linearineq}
\end{equation}
\begin{figure}
    \label{leggett}
	\includegraphics
	[width=0.5\textwidth]{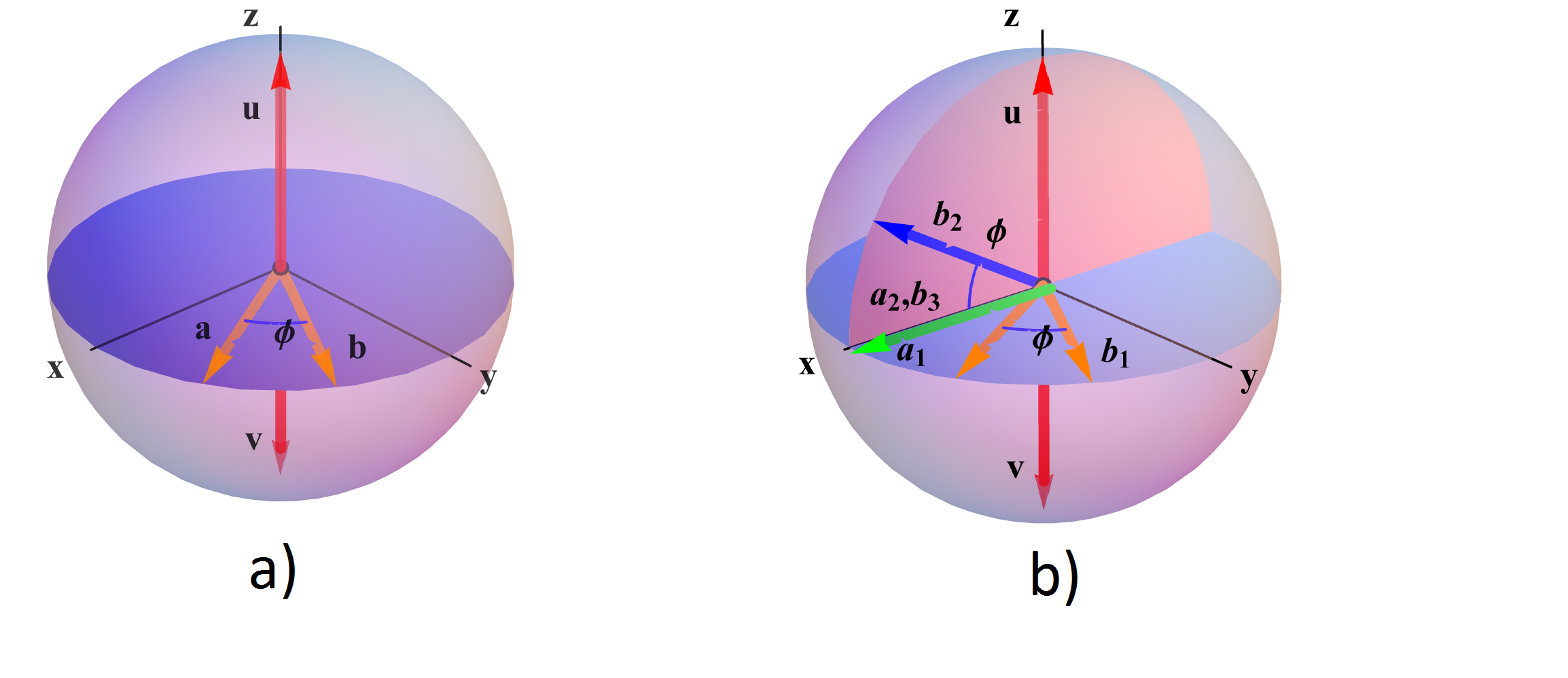}
	\caption{\textbf{a)} The nonlocal hidden variable model reproduces quantum correlations and can even violate the CHSH inequality for measurements performed along vectors \textbf{a} and \textbf{b} lying in the plane (purple) perpendicular to the \textbf{u} and \textbf{v}. \textbf{b)} Leggett's inequality is violated once the measurement vectors are not in the plane perpendicular to \textbf{u} and \textbf{v}.}
\end{figure}

For this hidden variables model to successfully give rise to quantum correlations, inequality~(\ref{linearineq}) needs to be satisfied. We observe that in a plane perpendicular to the initial polarization vectors $\mathbf{u}$ and $\mathbf{v}$, no matter what $\mathbf{a}$ and $\mathbf{b}$ we choose, the inequality~(\ref{linearineq}) is always satisfied. Hence for any measurement directions $\mathbf{a}$ and $\mathbf{b}$ lying in this plane (colored purple in Fig. 4), this model correctly predicts quantum correlations and hence we can also observe the violation of the CHSH inequality. However for some polarization vectors lying outside this plane, the inequality cannot be satisfied, and this is where the model fails.\newline

For a nonlocal hidden variable model as described above, when A uses two  different measurement settings $\mathbf{a_1}$, $\mathbf{a_2}$, and B uses three $\mathbf{b_1}$, $\mathbf{b_2}$ and $\mathbf{b_3} = \mathbf{a_2}$, a more general Leggett's inequality is given by,
\begin{eqnarray}
    S_\text{NLHV}&=&\abs{E_{11}(\phi)+E_{23}(0)} + \abs{E_{22}(\phi)+E_{23}(0)} \\ \nonumber
    &\leq& 4 - \frac{4}{\pi}\abs{\sin{\frac{\phi}{2}}},
\label{leggettineq}    
\end{eqnarray}
where, $E_{kl}$ is the expectation value of the product $\mathbf{A_k\,B_l}$ over all the initial polarization directions $\mathbf{u}$ and $\mathbf{v}$.

\begin{equation}
    E_{kl} = \int_{\mathbf{u},\mathbf{v}}\, d\mathbf{u}\, d\mathbf{v}\,F(\mathbf{u},\mathbf{v})\,\mathbf{A}(\mathbf{a_k},\mathbf{b_l},\lambda)\, \mathbf{B}(\mathbf{a_k},\mathbf{b_l},\lambda), 
\end{equation}
where $F(\mathbf{u},\mathbf{v})$ is the distribution of the initial polarization of two photons.
Quantum mechanically, this expectation value is given by,
\begin{equation}
    E_{kl} = -\mathbf{a_k} \cdot \mathbf{b_l}= -\cos{\phi_{\mathbf{a_k},\mathbf{b_l}}}.
\end{equation}
Substituting the expression in the left hand side of inequality~(\ref{leggettineq}), the quantity $S$ takes the value,
\begin{equation}
    S_{QM} = \abs{2(\cos{\phi}+1)}.
\end{equation}
For some values of $\phi$, $S_{QM} > S_\text{NLHV}$ goes above the upper bound set by Eq.~(\ref{leggettineq}) (the maximal violation occurs when $\phi = 18.8^{\circ}$). Hence, we can conclude that the crypto-nonlocal hidden variable theories also fail to fully describe quantum correlations.

Experiments performed with photon pairs entangled in polarization and spatial modes have indicated the violation of Leggett's inequality~\cite{nonlocalrealism1,nonlocalrealism2}.

\section*{Non-contextuality}
After looking into correlations between space-like separated quantum systems, we turn our focus to another feature of quantum mechanics, namely the context of the measurement.
Two physical observables are non-commuting if they do not have a set of simultaneous eigenstates. Therefore, the order in which one measures the two non-commuting quantities affects their measurement outcomes. Mathematically, the two measurement results, say for observables $\hat{A}$ and $\hat{B}$, are related by the uncertainty relation,
\begin{equation}
    \sigma_{A}\sigma_{B} \geq \abs{\frac{[\hat{A},\hat{B}]}{2i}},
\end{equation}
where $\sigma_{A}$ and $\sigma_{B}$ are the uncertainties in the two physical quantities $\hat{A}$ and $\hat{B}$. The commutator, $[\hat{A},\hat{B}]$, is zero if $\hat{A}$ and $\hat{B}$ commute with each other, i.e if their order of measurement does not matter, and is non-zero if they do not commute (i.e. if their order of measurement matters).
\[
  [\hat{A},\hat{B}] 
  \begin{cases}
                                  =0 & \text{if $\hat{A}$ and $\hat{B}$ commute.} \\
                                  \neq 0 & \text{if $\hat{A}$ and $\hat{B}$ do not commute.} 
  \end{cases}
\]

For the non-commuting physical quantities, we see that the uncertainty principle forbids simultaneous assignment of pre-determined measurement results. Moreover, if we now include a third observable $\hat{C}$ that also commutes with $\hat{A}$, i.e. $[\hat{A},\hat{C}]=0$, which does not need to commute with $\hat{B}$, i.e. $[\hat{B},\hat{C}]\neq 0$, the value assigned to $\hat{A}$ is considered to be non-contextual. In other words, the outcome of a measurement should not be different if the observable $\hat{A}$ is measured alone, together with $\hat{B}$ or together with $\hat{C}$. In the spirit of the EPR argument, one can now discuss non-contextual realism by assigning predefined values $v(\hat{A}_i)$ to all the observables $\hat{A}_i$.

 Therefore, non-contextuality, i.e. the notion that the measurement of a physical quantity is independent of the measurement of any other commuting physical quantities, or the ``context'' of the measurement, seemed like a valid assumption in quantum mechanics. However, Bell and Kochen-Specker (BKS)~\cite{Bellcontextuality,KochenSpeker,mermin} separately proved that it is impossible for the commuting observables to have pre-existing values independent of the context of the measurement. Kochen and Specker considered special kinds of observables that have binary eigenvalues (0 or 1), e.g. projection operators, and proved that it is impossible to assign values classically to the projection operators in a 3-dimensional Hilbert space. For the proof, they used projection operators along 117 different vectors. Cabello later provided a simpler proof~\cite{Cabello} of quantum contextuality, involving only 18 projection directions  in a four dimensional Hilbert space. Additionally, Klyachko, Can, Binicioglu and Shumovsky (KCBS) simplified it even further and found a proof that only requires 5 measurements for spin-1 particles~\cite{Klyachko}.

In the following, we focus on the latter, i.e. the KCBS version of the Bell-Kochen-Specker theorem, as it is the most simple BKS-proof regarding measurement settings and dimensionality of the quantum state~\cite{Klyachko,Cabello2}. Mathematically, if we assume quantum theory to be non-contextual, then for a state $\psi$ described by  commuting observables say $\{\hat{A}, \hat{B}, \ldots\}$, it is possible to assign an underlying value for the outcomes of each observables as say $\{v(\hat{A}), v(\hat{B}),\ldots\}$ independent and before the actual measurement (non-contextual realism). Since we know that the result of projective measurement of an observable can return only one of its eigenvalues, the value of the observable also must be one of the eigenvalues. 

Now classically if these observables satisfy the equation,
\begin{equation}
f(A, B, \ldots)=0,
\label{eq:function_obs}
\end{equation}
then the pre-assigned values should also satisfy the equation,
\begin{equation}
f(v(A), v(B), \ldots)=0 .
\label{eq:function_val}
\end{equation}
Note that a special case of this logical step, i.e. Eq.~(\ref{eq:function_obs}) $\Rightarrow$ Eq.~(\ref{eq:function_val}), is the so-called sum rule
\begin{equation}
A = B+C \Rightarrow v(A) = v(B) + v(C) .
\label{eq:sumrule}
\end{equation}

Let us now consider five numbers $a,b,c,d,e$ that can either take the value $+1$ or $-1$. For all possible combinations, the following algebraic inequality has a minimal value of $-3$: 
\begin{equation}
ab+bc+cd+de+ea \geq -3. 
\label{eq:inequality_val}
\end{equation}
This can be easily seen because at least one term always needs to be $+1$. As we already discussed above, according to a non-contextual hidden variable model each measurement has a predefined value, which is independent of the context, i.e. $v(A)_B = v(A)_C$. As this holds for all members of the ensemble, we can rewrite the inequality in terms of ensemble averages:
\begin{equation}
\left\langle AB\right\rangle+\left\langle BC\right\rangle+\left\langle CD\right\rangle+\left\langle DE\right\rangle+\left\langle EA\right\rangle \geq -3. 
\label{eq:inequality_ave}
\end{equation}
We note that this inequality holds not only for non-contextual hidden variable models but also for any joint probability distribution describing the measurements.   

However, KCBS realized that this inequality can be violated by quantum mechanics using five measurements of a spin-1 particle~\cite{Klyachko}. The measurements required are expressed by the spin operator $\hat{A}_i=2\hat{S}_i^2-I$, where the operator $\hat{S}_i^2$ has the eigenvalues 0 and 1 and, thus, needs to be rescaled and shifted to realize the required $\pm 1$ values. The five observables $\hat{A}_i$ are defined by the projection directions $\vec{l}_i$ according to:
\begin{equation}
\hat{A}_i=2\hat{S}_i^2-I=I-2\ket{\vec{l}_i}\bra{\vec{l}_i}.
\end{equation}
Two measurements $\hat{A}_i$ and $\hat{A}_j$ (for $i\neq j$) are commuting, i.e. are compatible, if and only if the projections $\vec{l}_i$ and $\vec{l}_j$ are orthogonal, which means that we need to find five pairwise orthogonal measurement directions. As the directions $\vec{l}_i$ can be directly depicted in a real three-dimensional space, we find that they form a pentagram. The maximal violation of the above described minimal value for non-contextual hidden variable can then be found for a quantum state $\Psi$ that lies on the symmetry axis of the pentagram (see Fig. \ref{pentagram}). If we now use these five pairwise orthogonal projections as our measurements on the quantum state $\Psi_0$, the quantum mechanical prediction is $5-4\sqrt{5}\approx -3.944$; thus, we can surpass the lowest limit given above. This simple proof demonstrates that quantum mechanics cannot be modeled by a non-contextual hidden variable model. Although there have been some discussions about the general validity of an experimental test of quantum contextuality \cite{meyer1999finite}, nowadays it has been generally accepted to be a valid task and many of experiments have been conducted~\cite{scarrino}, including one verifying the quantum mechanical prediction for the above given KCBS proof \cite{exptklyachko}.  

\begin{figure}
    \label{pentagram}
	\includegraphics[width=0.5\textwidth]{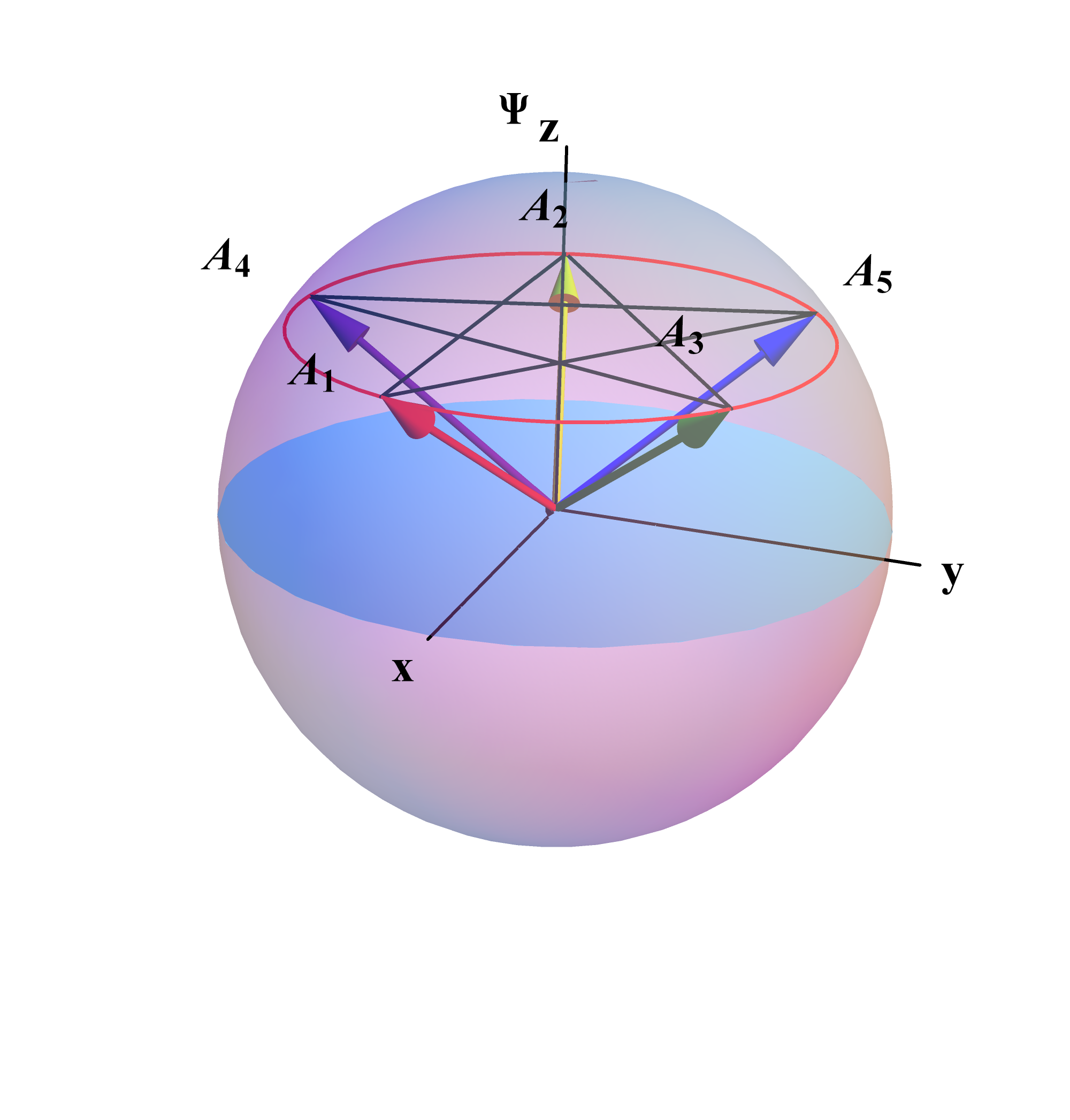}
	\caption{Three-dimensional representation of the measurement directions $\hat{A}_j$. We see that each measurement is orthogonal to two other directions and as such is compatible with it, i.e. commuting with these two other measurements. However, when using a quantum state $\Psi$ along the symmetry axis, the inequality for non-contextual hidden variable models given in formula (\ref{eq:inequality_ave}) can be violated.}
	\label{pentagram}
\end{figure} 

Interestingly, Bell searched for a more physical assumption than non-contextuality, and thus he studied locality instead. The important connection between the two arguments is that, simply phrased, in nonlocal hidden variable models, measurements cannot depend on measurements done in a space-like separated place. On the other hand, in non-contextual hidden variable models, measurements can ``only'' question the dependence of the measurement on the context if local or nonlocal. In other words, showing the mismatch between quantum mechanics and local realistic theories according to the Bell theorem is a stronger statement and as such includes (as a special case) non-contextual realistic theories. Thus, the possibility of space-like separation is an essential assumption in the Bell argument and requires two entangled particles that can be separated. 

\section*{Single particle ``Entanglement''}
So far we have only taken one degree of freedom of the quantum system into account. However, there are other quantum properties of single-particle systems that are similar in form to those of multipartite entangled states, i.e. non-separable correlations between two different degrees of freedom of a single particle. For instance, if we pass a diagonally polarized light through a polarizing beam splitter, it gets transformed to the state:
\begin{equation}
\ket{\psi} = \frac{1}{\sqrt{2}}\left(\ket{\text{I,H}} +\ket{\text{II,V}}\right),
\end{equation}
where the $\ket{\text{I,H}}$ indicates that the photon is in path $\text{I}$ and is horizontally polarized, and similarly for $\ket{\text{II,V}}$.
The mathematical expression for such states cannot be separated into a product of individual states in each of the two Hilbert spaces, quite similar to the entangled states. Although lacking nonlocality, these states have been shown to violate the CHSH inequality~\cite{singleparticleCHSH,stav:18}. Such ``entanglement'' tests can be used to probe the non-separability of such states. However, by using Bell-like inequalities, these experiments are challenging classical concepts that are closer to non-contextuality rather than nonlocality. Consequently, the violation of these inequalities rules out the ``local non-contextual hidden variable'' description for such states.\newline
We will discuss the details about such tests below~\cite{hardyspinorbit}, following an idea that was originally proposed for nonlocality by Hardy~\cite{HardyNonlocality}.
Let us look at three statements regarding probabilities, which, when simultaneously satisfied, would also imply a fourth statement in classical probability theory. We will then show that for certain quantum states and measurements in which the first three statements are true, the fourth statement turns out to be false, resulting in a contradiction with the classical picture.
We assume that:
\begin{equation}
\label{hardyprob}
\begin{array}{lcl}
\mathcal{P}(\alpha=+1,\beta =+1) = 0, \\
\mathcal{P}(\alpha=-1,\beta'=-1) = 0, \\
\mathcal{P}(\alpha'=-1,\beta=-1) = 0,
\end{array}
\end{equation}
where, $\alpha, \alpha', \beta$, and $\beta'$ are all binary variables which can take values $\pm1$. $P(\alpha=+1,\beta=+1)$ refers to the probability that $\alpha=+1$ and $\beta=+1$, and so on. Now if all three statements are true, then logically a fourth statement should follow:
\begin{equation}
\mathcal{P}( \alpha' =-1,\beta' =-1) = 0.
\end{equation}
Let us first see why this is true for a classical non-contextual theory. We tabulate all the possibilities for these binary variables in Table III \ref{TableIII}.
\begin{table}[ht]
    \label{TableIII}
	\centering
	\caption{\textbf{Possible combinations of the values of the binary variables $\mathbf{\alpha}$,
		$\mathbf{\beta}$,
		$\mathbf{\alpha'}$, and
		$\mathbf{\beta'}$ as per the classical non-contextual theory. } $\mathcal{P}_{i}$s are the corresponding probabilities for the particular combination of binary values of the variables.}
	\begin{tabular}{@{} |p{2.0cm}|p{1.5cm}|p{1.5cm}|p{1.5cm}|p{1.5cm}| @{}} \hline
		\textbf{Probability} & $\mathbf{\alpha}$ &	$\mathbf{\beta}$ &	$\mathbf{\alpha'}$ & $\mathbf{\beta'}$ \\ \hline
		$\mathcal{P}_1$ & $+1$ & $+1$ & $+1$ & $+1$ \\ \hline
		$\mathcal{P}_2$ & $+1$ & $+1$ & $+1$ & $-1$ \\ \hline
		$\mathcal{P}_3$ & $+1$ & $+1$ & $-1$ & $+1$ \\ \hline
		$\mathcal{P}_4$ & $+1$ & $+1$ & $-1$ & $-1$ \\ \hline
		$\mathcal{P}_5$ & $+1$ & $-1$ & $+1$ & $+1$ \\ \hline
		$\mathcal{P}_6$ & $+1$ & $-1$ & $+1$ & $-1$ \\ \hline
		$\mathcal{P}_7$ & $+1$ & $-1$ & $-1$ & $+1$ \\ \hline
		$\mathcal{P}_8$ & $+1$ & $-1$ & $-1$ & $-1$ \\ \hline
		$\mathcal{P}_9$ & $-1$ & $+1$ & $+1$ & $+1$ \\ \hline
		$\mathcal{P}_{10}$ & $-1$ & $+1$ & $+1$ & $-1$ \\ \hline
		$\mathcal{P}_{11}$ & $-1$ & $+1$ & $-1$ & $+1$ \\ \hline
		$\mathcal{P}_{12}$ & $-1$ & $+1$ & $-1$ & $-1$ \\ \hline
		$\mathcal{P}_{13}$ & $-1$ & $-1$ & $+1$ & $+1$ \\ \hline
		$\mathcal{P}_{14}$ & $-1$ & $-1$ & $+1$ & $-1$ \\ \hline
		$\mathcal{P}_{15}$ & $-1$ & $-1$ & $-1$ & $+1$ \\ \hline
		$\mathcal{P}_{16}$ & $-1$ & $-1$ & $-1$ & $-1$ \\ \hline
	\end{tabular}
	\label{table3}
\end{table}

Looking at Table III\ref{TableIII}, the first statement, i.e.  $P(\alpha=+1,\beta =+1) = 0$ implies,
\begin{align}
   \mathcal{P}_1 + \mathcal{P}_2 + \mathcal{P}_3 +\mathcal{P}_4 &= 0.
\end{align}
Similarly, from the second and third statements,
\begin{align}
    \mathcal{P}_{10} + \mathcal{P}_{12} + \mathcal{P}_{14} + \mathcal{P}_{16} &= 0 \\
    \mathcal{P}_7 + \mathcal{P}_8 + \mathcal{P}_{15} + \mathcal{P}_{16} &= 0.
\end{align}
Since probabilities cannot be negative, these statements imply that the individual probabilities $P_1, P_2, P_3, P_4, P_7, P_8, P_{10}, P_{12}, P_{14}, P_{15}, P_{16}$ are all zero. Now, for the fourth statement to be true, the following expression must be satisfied,
\begin{align}
    \mathcal{P}_4 + \mathcal{P}_8 + \mathcal{P}_{12} + \mathcal{P}_{16} = 0.
\end{align}
We observe that all the individual terms on the left hand side are already zero from the first three statements. Hence, it follows that the fourth statement must be true.

Now in the quantum version, we assume that our usual two parties, namely Alice and Bob, each have access to one of the two different degrees of freedom of the single particle quantum state:
\begin{equation}
\ket{\psi} = \cos\gamma \ket{0}_{A}\ket{1}_{B}-\sin\gamma\ket{1}_{A}\ket{0}_{B}.
\end{equation}
The two kets can represent any two degrees of freedom of a single particle (for e.g. path and polarization) in the two dimensional Hilbert space. Alice can perform a measurement of two projection operators $\alpha $ and $\alpha'$,
and Bob can perform his measurement of projection operators $\beta$ and $\beta'$.

The projection operators $\alpha$, $\beta$ and $\alpha'$, $\beta'$ are defined as,
\begin{eqnarray}
\alpha, \beta &=&
\ket{+}\bra{-}-\ket{-}\bra{+}
\\
\alpha', \beta' &=&
|+\rangle'\langle-|-|-\rangle'\langle+|.
\end{eqnarray}
where,
\begin{eqnarray}
\ket{+} &=& N\left(\sqrt{\sin\gamma}|0\rangle+\sqrt{\cos\gamma}\ket{1}\right)
\\
\ket{-} &=& N\left(-\sqrt{\cos\gamma}|0\rangle+\sqrt{\sin\gamma}\ket{1}\right)
\\
\ket{+}' &=& N'\left(\sqrt{{\cos^3\gamma}}|0\rangle+\sqrt{{\sin^3\gamma}}\ket{1}\right)
\\
\ket{-}' &=& N'\left(-\sqrt{{\sin^3\gamma}}|0\rangle+\sqrt{{\cos^3\gamma}}\ket{1}\right)	
\end{eqnarray}
and $ N = \frac{1}{\sqrt{\sin\gamma+\cos\gamma}}$ and $N' =\frac{1}{ \sqrt{\sin^{3}\gamma+\cos^{3}\gamma}}$ are the normalization constants.
For these states we find that the probabilities are as defined in Eq.~(\ref{hardyprob}). However the expression for the fourth probability is,
\begin{equation}
P(\alpha'=-1,\beta'=-1) = \left( \frac{\sin 4\gamma}{4(\cos^3\gamma+\sin^3\gamma)}\right)^2
\end{equation}
For certain angles $\gamma$, this expression is not zero which is in conflict with the assumption that the values of the four operators can pre-exist before we conduct the measurement. Thus, it disproves the non-contextual hidden variable description for non-maximally separable states.

Spin-energy entangled states in massive single particles such as neutrons have also been used~\cite{hasegawa} to demonstrate the violation of Leggett type inequalities for the contextual realistic hidden variables theories.

\section{N00N states}
After having discussed various scenarios where quantum systems involving one or two particles, show features that cannot be explained with classical theories, we now turn to many-body quantum systems that resemble bipartite systems and exhibit interesting correlations. One such state is a N00N state \cite{BarryNoon,BotoNoon,LeeNoon}, which is an equal superposition of $N$ indistinguishable particles in one mode and none in the other, and vice versa. Mathematically, such states have the form,
\begin{equation}
    \ket{\psi} = \frac{1}{\sqrt{2}}\Big(\ket{N}_{a}\ket{0}_{b} +  \ket{0}_a\ket{N}_b\Big),
\end{equation}
where the subscripts ``a'' and ``b'' now refer to the modes a and b which photons can occupy, in contrast to our previous notation where the subscripts represented the photons themselves. These states are sometimes also referred to as Schr\"odinger cat states \cite{DowlingNoon} as they represent a superposition of $N$ particles (in theory $N$ could be arbitrarily large) in two distinct states, comparable to the ``dead'' or ``alive'' states. These N00N states have interesting applications in quantum metrology \cite{DowlingNoon,JonesNoon,IsraelNoon} and can provide sensitivity even upto to the Heisenberg limit.

For our purposes, N00N states present an interesting case of ``entanglement''. For instance, let us consider the simplest N00N state with just a single particle,

\begin{equation}
    \ket{\psi} = \frac{1}{\sqrt{2}}\Big(\ket{1}_{a}\ket{0}_{b} +  \ket{0}_a\ket{1}_b\Big).
\end{equation}

Expressed in the number basis, here it looks like we have a single particle ``entangled'' with the vacuum. It gets even more intriguing when we have a particle in this state interacting with other atoms or particles. Let us look at a photon in such a state, which is generated by passing a photon through a $50$:$50$ beamsplitter, and creating a superposition between the two different paths (modes). In each of the two paths, we place an atom in a ground state such that it jumps into an excited state when it comes in contact with the photon. Unless we perform a measurement, we do not know which path the photon has taken and which one of the atoms is in the  excited state. At this instant, the joint state of the two atoms is,

\begin{equation}
    \ket{\psi} = \frac{1}{\sqrt{2}}\Big(\ket{e}_{a}\ket{g}_{b} +  \ket{g}_a\ket{e}_b\Big),
\end{equation}
where $\ket{e}_{a}\ket{g}_{b}$ denotes that the atom in mode a is in the excited state and the atom in mode b is in the ground state, and so on.
Now this is unequivocally an entangled state. As any local operations cannot increase or decrease entanglement, this also strengthens the claim that the single photon state posses some characteristics similar to an entangled state.
Similar arguments on the ``nonlocality'' of a single particle have been much debated in the literature~\cite{TanSinglephoton,HardySinglephoton,GreenbergerSinglephoton,VaidmanSinglephoton,AharonovSinglephoton}.

N00N states with two indistinguishable particles also offer an interesting case. For instance, let us look at two such photons in a 2002 state, with equal superposition of either both being diagonally polarized or anti-diagonally polarized,
\begin{equation}
 \ket{\psi} = \frac{1}{\sqrt{2}}\Big(\ket{2}_{A}\ket{0}_{D} -  \ket{0}_A\ket{2}_D\Big). 
\end{equation}
When written in a terms of individual kets of each photons,
\begin{equation}
 \ket{\psi} = \frac{1}{\sqrt{2}}\Big(\ket{A}_{1}\ket{A}_{2} -  \ket{D}_1\ket{D}_2\Big). 
\end{equation}

This two photon state can be obtained by passing two indistinguishable photons in a state,
\begin{equation}
    \ket{\psi} = \ket{1}_H\ket{1}_V \label{sec_quant_state}
\end{equation}
through a polarizing beamsplitter (PBS) set at an angle of $45^\circ$,
\begin{eqnarray*}
    \ket{\psi}&=& \hat{a}^\dag_H \hat{a}^\dag_V  \ket{0} \\
              &=& \frac{1}{2}(\hat{a}^\dag_A + \hat{a}^\dag_D)(\hat{a}^\dag_A-\hat{a}^\dag_D) \ket{0} \\
              &=& \frac{1}{2}(\hat{a}^\dag_A \hat{a}^\dag_A + \hat{a}^\dag_A \hat{a}^\dag_D - \hat{a}^\dag_D \hat{a}^\dag_A - \hat{a}^\dag_D \hat{a}^\dag_D)\ket{0}\\
              &=& \frac{1}{2}(\hat{a}^\dag_A \hat{a}^\dag_A- \hat{a}^\dag_D \hat{a}^\dag_D)\ket{0}\\
              &=&\frac{1}{\sqrt{2}}\Big(\ket{2}_{A}\ket{0}_{D} -  \ket{0}_A\ket{2}_D\Big).
\end{eqnarray*}
The two photons are clearly entangled with each other after passing through the beamsplitter. Note that this is the famous Hong-Ou-Mandel (HOM) interference for two identical photons passing through a beamsplitter \cite{hongoumandel}. One crucial point to be made here is that the entangled state is created by the physical action of the beamsplitter on both of these photons. It may seem that the entangled state could be obtained simply by just changing the basis of polarization as,
\begin{eqnarray}
    \ket{\psi} &=&  \ket{H}_1\ket{V}_2 \label{wrong_state}\\
               &=&\frac{1}{2}(\ket{A}_1+\ket{D}_1)(\ket{A}_2-\ket{D}_2 )\nonumber\\
               &=& \frac{1}{2}(\ket{A}_1\ket{A}_2-\ket{A}_1\ket{D}_2+\ket{D}_1\ket{A}_2-\ket{D}_1\ket{D}_2). \notag
\end{eqnarray}
and then canceling the two terms in the middle. However, there are two things that have to be considered. First, since we are dealing with two indistinguishable photons, it is essential to symmetrize the initial state and hence Eq.~(\ref{wrong_state}) does not characterize the state of two indistinguishable photons. The second state, written in the number basis (in Fock space), i.e. Eq.~(\ref{sec_quant_state}), is the correct approach for writing such states. It captures  all the possible combination of the two photons in such a state. Another fundamental reason is that the physics, or the phenomenon of entanglement in this case, should not change just by altering the basis. In contrast to the discussion before, where we had a beamsplitter performing a joint physical action on the two photons, here only a rotation of the coordinate system has been performed. Such a rotation cannot lead to a change or creation of entanglement.

\section{Bounds on quantum correlations}
In the preceding sections, we saw how local, crypto-nonlocal or non-contextual hidden variable theories are bounded by Bell-like inequalities and how quantum correlations can violate these and reach beyond the classical bounds. We now shift our focus to the other side of this picture, i.e. \emph{to what extent quantum mechanics departs from classical physics and what promise such maximal violations hold for unique quantum technologies?} Mathematically, the maximal violation of a Bell inequality is given by the Tsirelson bound, following the seminal work in this respect \cite{tsirelsonbound}. It was shown that,
\begin{equation}
 S = \langle A_0\,B_0\rangle + \langle A_0\,B_1\rangle+\langle A_1\,B_0\rangle - \langle A_1\,B_1\rangle \le 2\sqrt{2},
\end{equation}
where $\langle A_i\,B_j \rangle$ are the expectation values of the product of Alice's and Bob's $\pm 1$-valued observables $A_i$ and $B_j$, respectively. Quantum correlations can therefore violate the CHSH inequality \cite{CHSHineq}, $S\le 2$ mentioned above by at most a multiplicative factor of $\sqrt{2}$. 

Despite its importance, the Tsirelson bound provides only the point of maximal violation; hence, a much more detailed characterization of quantum mechanics would consist of all nonlocal correlations achievable by quantum operators acting on quantum states, i.e. the quantum set of correlations. In between these two descriptions, there are partial characterizations of quantum correlations such as the Uffink inequality \cite{uffnik}, some of which have already been tested experimentally with a high-fidelity source of polarization-entangled photons \cite{limitsofnonlocality}. However, a general, finite characterization of the quantum set is still missing. In the simplest bipartite case with binary inputs and outputs, the TLM (Tsirelson-Landau-Masanes) inequality is known to be necessary and sufficient for the correlators to be realizable in quantum mechanics \cite{T,L,M} 
\begin{equation}
|c_{00}\,c_{10}-c_{01}\,c_{11}|\le \sum_{j=0,1}\sqrt{(1-c_{0j}^2)\,(1-c_{1j}^2)},
\end{equation}
where we have used $c_{ij}=\langle A_iB_j \rangle$. In recent years, there has been a growing interest in further exploring this set of quantum correlations from within and from outside the quantum formalism ~\cite{EliPR,EliNPA,EliNew,Nlbeyondqm,Geometry,boundnetworks} in order to derive the strength of quantum correlations based on first principles~\cite{EliIC,EliUnc,EliML,EliLO,EliCC,NLcomputation}. 

According to a recent approach \cite{EliCarmi1}, some well-known bounds on quantum correlations (such as the Tsirelson and TLM bounds), as well as some new ones, originate from a principle called ``relativistic independence'', encapsulating relativistic causality and indeterminism. This means that even a very general probabilistic structure can give rise to quantum-like correlations (but not stronger-than-quantum correlations) if it obeys a generalized uncertainty principle (i.e. an uncertainty principle which is applicable even beyond quantum mechanics, but when assuming an Hilbert space structure reduces to the well-known Schr\"{o}dinger-Robertson uncertainty relation), which is moreover local. Locality in this context means that choices made by remote parties do not affect the uncertainty relations of other parties. This result quantitatively supports a famous conjecture~\cite{EliConS1,EliConS2,EliConA}  arguing that quantum mechanics can be as nonlocal as it is without violating relativistic causality only due to its inherent indeterminism. Moreover, it shows that entanglement-assisted nonlocal correlations and uncertainty are two aspects of the same phenomenon, imprinted in the algebra of quantum mechanics (which can be accessed from outside the quantum formalism using local uncertainty relations). This result therefore attributes the differences between classical and quantum correlations to the existence of fundamental uncertainty within quantum mechanics. Several applications of this approach can be found in~\cite{EliCarmi2,EliCarmi3,EliCarmi4}. For related analyses see \cite{Rel1,Rel2}.

\section*{``Classical Entanglement''}
Having discussed different aspects of quantum physics, such as realism, locality and non-contextuality for single and multiple particles, we can now study how concepts such as ``classical entanglement'' relate to these fundamental concepts. Some works~\cite{spreeuw,oxymoron,eberly:2015,classicalbellandqft,simbellwithclassical,shiftquant,classicalentpolmet} have compared the true quantum phenomenon of entanglement with  classical waves of light and called the analogies ``classical entanglement''. For instance, instead of single photons if we send a classical electromagnetic wave through the polarizing beamsplitter, then the electric field is now written as a superposition of the two components,
\begin{equation*}
	\mathbf{E}(\mathbf r) = E_{H}(\mathbf r)\,\mathbf{e}_{H} + E_{V}(\mathbf{r})\,\mathbf{e}_{V},
\end{equation*}
where $\mathbf{e}_{H}$ and $\mathbf{e}_{V}$ are unit vectors along horizontal and vertical directions, and $E_{H}(\mathbf r)$ and $E_{V}(\mathbf r)$ are the corresponding electric field components, respectively. The electric field and the polarization are correlated and the intensities also violate inequalities that resemble Bell inequalities~\cite{singleparticleCHSH,belllikeborges,saleh:2013}. However, in this case, classical fields are used instead of single particles, such as photons, and, therefore, we are not performing tests of the assumptions such as realism, locality, non-contextuality or any class of  hidden variables. As all these considerations using classical states of light, i.e. coherent states, are fully described by Maxwell's equations, i.e. only require a wave picture without invoking the field quantization, they cannot challenge any of the above mentioned fundamental concepts. All contradictions to classical concepts and mind-boggling questions arose upon considering the \emph{particle} nature of light, i.e. when using single photons. Hence, it is misleading to challenge fundamental concepts using states of light that are fully described by the electromagnetic wave picture and Maxwell's equations. Therefore, we suggest that the term entanglement should only be used in connection to quantum experiments with single or multiple particles, and in particular for the cases involving non-locality as it was originally suggested by Schr\"{o}dinger. Using the equally valid term of ``non-separability'', which is not as closely related to fundamental ideas as entanglement, might be more appropriate and simplify the distinction between classical and quantum correlations, for experts as well as the interested layman. Moreover, although analogies might be correct, the beauty as well as deep implications due to quantum entanglement might otherwise be misinterpreted, oversimplified or even entirely misunderstood. 
This type of classical states, i.e. non-separable states, nevertheless have important applications~\cite{oxymoron} for example in polarization metrology~\cite{classicalentpolmet}, kinematic sensing~\cite{kinematictracking}, computation~\cite{classicalcomputation}, communication~\cite{Li:2016,andrew:2017}, and many more~\cite{localteleportation,gerd:2019}.

\section{Conclusion}
 Stemming directly from the principles of quantum mechanics, the presence of entanglement in any multipartite system marks a distinct departure from classical physics. Defying any classical explanation, entanglement raises some intriguing fundamental questions about the physical universe such as realism, non-locality, etc. Some exciting manifestations of entanglement, such as  Schr\"odinger cat states and N00N states, serve to question the boundary between the quantum and classical world, highlighting some stark differences between the two regimes.
 Non-separable single-particle states, similar to entangled states in their mathematical form, yet lacking non-locality, present some intriguing instances of correlations allowing one to investigate the meaning of contextuality. Classical electromagnetic phenomena analogous in their form to entangled states, although useful in a variety of applications~\cite{classicalentpolmet,kinematictracking,classicalcomputation,Li:2016,andrew:2017,localteleportation, gerd:2019}, cannot act as tests of the fundamental concepts of non-locality, realism or contextuality as entangled states can. As such, we suggest ``non-seperability'' as a more appropriate term for these states to clearly distinguish them from the quantum phenomenon of entanglement.

 In addition to being an integral concept of quantum foundations, entanglement is also a  key resource in modern technological advances in quantum computing, quantum communication and quantum metrology. To a large extent, the second quantum revolution we are witnessing these days strongly relies on generation and manipulation of entangled quantum states. Bell inequalities~\cite{bell} and Tsirelson bounds~\cite{tsirelsonbound} quantify the lower and upper limits, respectively, on the correlations obtainable from entangled states to be non-classical and still considered quantum. Striking forms of quantum correlations, different from entanglement, have also been studied, most notably, quantum discord~\cite{discord1,discord2,discord3,discord4}, which poses some avenues for future research. Moreover, nonlocality itself is believed to be a broader phenomenon than presented here, often including dynamical nonlocality~\cite{dynamicnonlocality,finallymakingsense}, such as the one commonly attributed to the Aharonov-Bohm effect~\cite{aharonovbohm1,aharonovbohm2}, which is also closely connected to entanglement \cite{EAB1,EAB2,EAB3,EAB4}.
 This type of nonlocality still merits further quantitative study~\cite{dynamicnonlocality}.  

\section{Acknowledgements}
We thank Leon Bello and Avishy Carmi, and members of the SQO group for helpful comments. This work was supported by Canada Research Chairs (CRC), Canada First Excellence Research Fund (CFREF), and Ontario's Early Researcher Award. RF acknowledge the support of the Academy of Finland through the Competitive Funding to Strengthen University Research Profiles (301820) and the Photonics Research and Innovation Flagship (PREIN - 320165).


\begin{thebibliography}{999}
\bibitem{photoelektric}
    Actually, the photoelectric effect can also be described, to some extent, by classical electrodynamics but quantum mechanics offers far more clear and intuitive explanation.
\end{thebibliography}
\end{document}